\documentclass[pdflatex,sn-mathphys-num]{sn-jnl}

\usepackage{graphicx}%
\usepackage{multirow}%
\usepackage{amsmath,amssymb,amsfonts}%
\usepackage{amsthm}%
\usepackage{mathrsfs}%
\usepackage[title]{appendix}%
\usepackage{xcolor}%
\usepackage{textcomp}%
\usepackage{manyfoot}%
\usepackage{booktabs}%
\usepackage{algorithm}%
\usepackage{algorithmicx}%
\usepackage{algpseudocode}%
\usepackage{listings}%

\theoremstyle{thmstyleone}%
\theoremstyle{thmstyletwo}%

\theoremstyle{thmstylethree}%

\begin{document}

\title{Effect of Synchrotron Radiation on Staged Plasma Wakefield Accelerators}


\author*[1]{\fnm{Livio} \sur{Verra}}\email{livio.verra@lnf.infn.it}

\author[2]{\fnm{Alexander} \sur{Knetsch}}

\author[2]{\fnm{Doug} \sur{Storey}}

\affil*[1]{\orgdiv{INFN}, \orgname{Laboratori Nazionali di Frascati}, \orgaddress{\street{Via Enrico Fermi 54}, \city{Frascati}, \postcode{00044}, \country{Italy}}}

\affil[2]{\orgdiv{SLAC National Accelerator Laboratory}, \orgaddress{\street{2575 Sand Hill Rd}, \city{Menlo Park}, \postcode{94025}, \state{CA}, \country{USA}}}


\abstract{
In a staged, beam-driven, plasma wakefield accelerator, electrons are accelerated in a sequence of plasma stages, each powered by a new driver electron bunch.
Between each stage, a magnetic chicane is used to dispose of the spent driver and to inject the new fresh one, while transporting the witness bunch. 
We discuss the effect of synchrotron radiation in the interstage sections on the accelerating gradient and on the final energy reach of such a machine.
}

\maketitle

\section{Introduction}\label{sec1}
Plasma wakefield acceleration is one of the promising techniques to deliver particles to a high-energy collider at the energy frontier at an affordable cost, exploiting the extremely high accelerating gradients that can be generated in plasmas~\cite{DAWSON:1959,TAJIMA:1979,CHEN:1985}.

\par In the beam-driven plasma wakefield acceleration (PWFA) configuration, a charged particle bunch (the $driver$) propagating in plasma drives a plasma electron density perturbation that sustains longitudinal and transverse fields (the $wakefields$).
A trailing lower-charge bunch (the $witness$) traveling in the appropriate phase of the wakefields (accelerating and focusing) can be accelerated with high gradient and efficiency~\cite{LITOS:2014}. 
In practice, the plasma acts as a transformer, transferring energy from the higher-charge driver to the lower-charge witness bunch~\cite{CHEN:1986}.

\par In the context of the 10 TeV Wakefield Collider Design Study~\cite{GESSNER:2025}, we investigate the feasibility of accelerating electrons ($e^-)$ to $5\,$TeV with a PWFA linac for $e^-e^-$ or photon ($\gamma \gamma$) collisions~\cite{BARKLOW:2023,BARKLOW:2023b}.
A wakefield-based collider could be an energy upgrade of an existing linear collider facility (LCF)~\cite{LCVISION:2025}. 
Considering the length of one linac arm of LCF ($\sim 10\,$km), the target effective gradient to reach 5\,TeV is $W_\mathrm{eff}=E/Z=0.5\,$GV/m ($E$ the final beam energy, $Z$ the length of the linac).

\par There are two ways to reach the target energy of 5\,TeV:
one is to use one high-energy driver that drives wakefields and accelerates the witness in a single, very long, plasma stage;
the other option is to accelerate the witness in a series of shorter plasma stages, each powered by a lower energy (hence, cheaper~\cite{FOSTER:2025}) driver bunch. 

\par In the first case, an extremely high energy driver bunch is needed, such as proton bunches from synchrotrons, to drive large-amplitude wakefields over long distances. 
Even though proton-driven PWFA in a single stage has been demonstrated~\cite{AWAKE:2018}, the acceleration process relies on the self-modulation instability~\cite{KUMAR:2010}, since the driver bunch length is orders of magnitudes longer than the plasma electron wavelength. 
This process can be controlled with appropriate seeding methods~\cite{BATSCH:2021,VERRA:2022b}, but it may prevent acceleration of bunches with the extremely high quality required for high-luminosity colliders. 
One solution would be to compress the proton bunch before injection in plasma~\cite{CALDWELL:2009}, but substantial R$\&$D is required to demonstrate the feasibility of the scheme~\cite{CALDWELL:2025}. 

\par In the case of a staged PWFA linac~\cite{LINDSTROM:2021b}, lower-energy, independent driver electron bunches can be used to drive the wakefields in each plasma stage, whereas the witness bunch is transported from one stage to the next, while preserving (or even improving~\cite{LINDSTROM:2021c}) its transverse and longitudinal quality. 
In the following, we investigate the effect of synchrotron radiation in the interstage sections on the accelerating gradient and on the energy reach of such a machine. 
\begin{figure}[!h]
\centering
\includegraphics[width=\columnwidth]{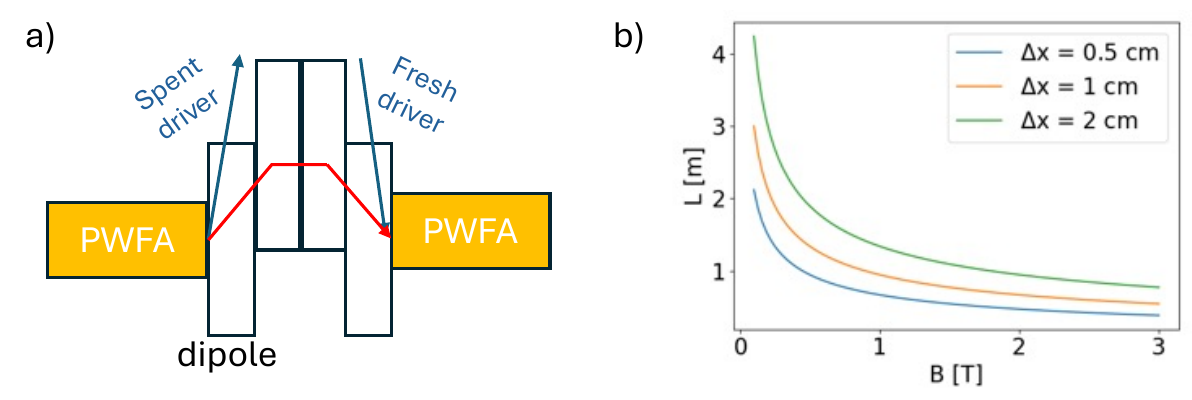}
\caption{(a): Schematic of the simplest period of a staged PWFA linac, composed by the plasma stage, a chicane of four dipoles to separate the witness bunch (red arrow) from the spent driver bunch (blue arrow on the left) and to inject the fresh driver (blue arrow on the right) into the following plasma stage;
(b): Required length of a magnetic dipole to separate by $\Delta x$ a witness bunch with $E_W=27\,$GeV from a driver bunch with maximum energy $E_D=9\,$GeV, as a function of the magnetic field amplitude $B$. }
\label{fig:1}
\end{figure}
\section{Staging}\label{sec2}

The basic period of a staged PWFA consists of (see Fig.~\ref{fig:1}(a)): the plasma stage, a dipole to separate in the dispersive plane the spent driver from the witness, a dipole to inject the fresh new driver and the witness in the following plasma stage, and two dipoles in between to bring the witness back on the original propagation axis, cancel the transverse dispersion, and control the transport matrix element $R_{56}$ of the chicane~\cite{DIMITRI:2016,LINDSTROM:2025}. 

\par In general, the interstage sections should be as short as possible to maximize the effective accelerating gradient of the linac.
The length of the first dipole is determined by the required transverse separation $\Delta x$ in the dispersive plane between driver and witness bunches, and by the amplitude of the magnetic field $B$, according to~\cite{LINDSTROM:2021b}:
\begin{equation}
    \label{eq:1}
    \Delta x=\frac{BqcL^2}{2} \left( \frac{1}{E_W} - \frac{1}{E_D} \right),
\end{equation}
where $q$ is the elementary charge, $L$ is the length of the dipole, and $E_D$ and $E_W$ are the energies of driver and witness bunches, respectively, at the exit of the prior plasma stage.
In the following, we conservatively consider $E_D$ equal to the initial energy of the driver electrons, to take into account that some electrons in the front of the bunch may lose only a negligible amount of energy. 
Moreover, an identical dipole at the end of the chicane can then be used to inject the fresh new driver (coming with the same offset $\Delta x$ into the following stage.)
Figure~\ref{fig:3}(b) shows that the required length of the dipole decreases when increasing $B$, and for shorter $\Delta x$.
In the case of positron witness bunches, the particles of the two beams are bent in opposite directions, and the two terms in parenthesis in Eq.~\ref{eq:1} are added rather than subtracted, yielding a larger $\Delta x$.

\par The minimum $\Delta x$ is defined by the dumping system for the depleted driver bunch. 
For example, the driver could be captured by the magnetic field of a septum magnet, leaving the witness bunch unperturbed~\cite{Barnes:2010}.

\par As we discuss later, a realistic period would also include optical elements to properly transport and match the witness beam envelope from one plasma stage to the next one.

\section{Effect of Synchrotron Radiation}

As a first approximation, we consider the chicane as made of four identical dipoles, with no empty space between magnets and the plasma stages. 
We also consider $E_D$ and initial energy of the witness $E_W=9\,$GeV, in-plasma accelerating gradient $W_z^+=5\,$GV/m and 3.6-m-long plasma stages (i.e., energy gain $\Delta E^+=18\,$GeV per stage), and a transformer ratio $R=W_z^+/W_z^-=2$ ($W_z^-$ the maximum decelerating field within the driver bunch).
These are reasonable assumptions for state-of-the-art, high-quality and high-energy PWFA~\cite{LITOS:2014, ZHANG:2025}.

\par Since the witness bunch is forced on a curved trajectory by the dipoles in each chicane, it emits synchrotron radiation~\cite{JACKSON:1962}.
The average energy loss in eV units of an electron through a dipole is calculated classically as: 
\begin{equation}
    U_0 = \frac{q}{6\pi\varepsilon_0}\beta^3 \gamma^4 L \left(\frac{B}{3.3pc}\right)^2,
\end{equation}
where $p$ is longitudinal momentum of the witness at the plasma exit (in eV/c units), $\gamma$ the Lorentz factor and $\beta$ is the beam velocity normalized to the speed of light $c$ and $\varepsilon_0$ is the vacuum permittivity.

\par
\begin{figure}[!h]
\centering
\includegraphics[width=\columnwidth]{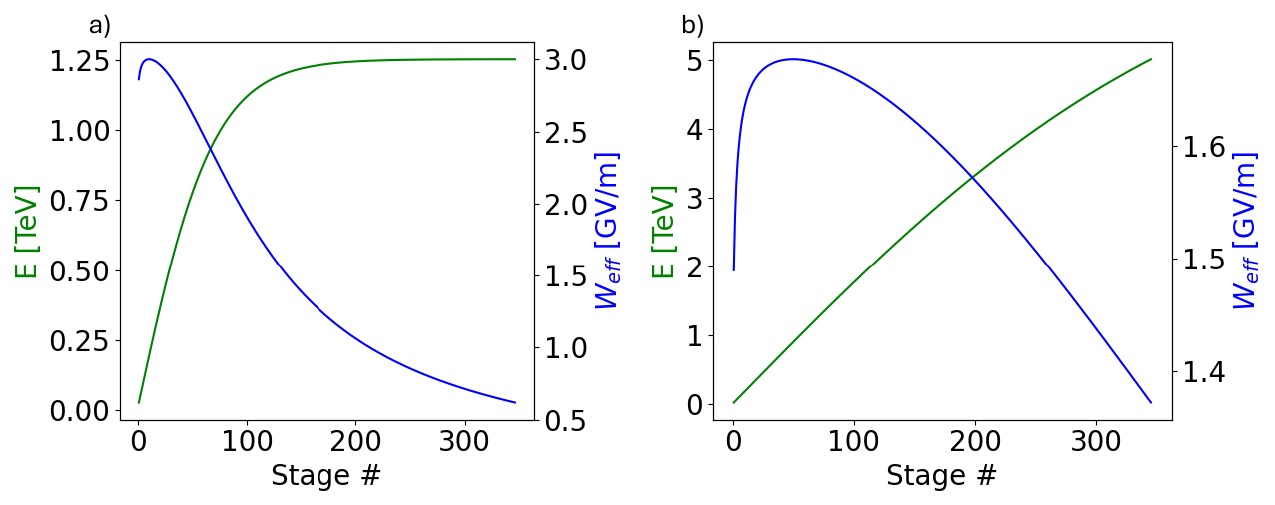}
\caption{Energy of the witness bunch (green line, left-hand side vertical axis) and effective gradient (blue line, right-hand side axis vertical axis), along the PWFA linac for 2-T (a) and 0.2-T (b) dipoles.}
\label{fig:2}
\end{figure}

 Figure~\ref{fig:2}(a) shows that, for the case of 2-T dipoles with appropriate length to achieve $\Delta x=0.5\,$cm, the energy of the witness bunch (green line) increases along the linac but saturates around 1.25\,TeV. 
This is due to the fact that, as the energy of the witness increases, so does the average energy loss in each interstage section. 
Therefore the effective $W_\mathrm{eff}(z)=E_W(z)/z$ ($z$ the propagation distance along the linac) starts decreasing, and the energy saturates when the energy loss in each interstage section becomes comparable to the energy gain in the following plasma stage, preventing a further increase of the witness bunch energy. 
\par A solution to achieve the target energy of 5\,TeV is to decrease the amplitude of the magnetic field in the dipoles [Fig.~\ref{fig:2}(b) shows the case with $B=0.2\,$T], at the expenses of having $W_\mathrm{eff}$ considerably lower than $W_z^+$ (5\,GV/m in this case).

\par The basic interstage section we have presented would naturally be complicated in a real machine by the presence of optical elements to capture and re-inject the witness from one plasma stage to the following one. 
When the witness beam is transversely matched to the plasma ion column focusing force, it leaves the plasma at a waist with the matched Twiss $\beta$ function $\beta_m=\sqrt{2\gamma}c/\omega_{pe}$~\cite{MUGGLI:2004,VERRA:2020}, where $\omega_{pe}=\sqrt{n_{pe}q^2/m_e\varepsilon_0}$ the plasma electron angular frequency ($n_{pe}$ the plasma electron density, $m_e$ the electron mass).
Since $n_{pe}$ is assumed to be the same in all plasma stages, the beam must be re-injected in plasma with the same matched parameters ($\alpha=0$, $\beta=\beta_m$). 
Thus, a focusing system is required (drifts plus active elements) to reduce the divergence of the witness beam at the plasma exit, and to re-inject and match it at the entrance of the following stage with $1:1$ magnification. 


\par
\begin{figure}[!h]
\centering
\includegraphics[width=\columnwidth]{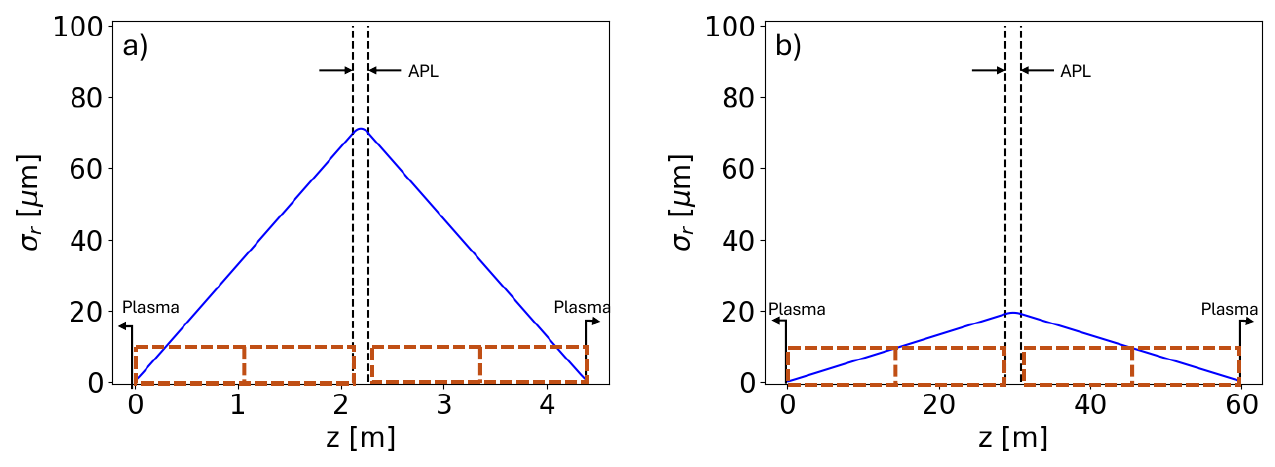}
\caption{Root-mean-square of the transverse distribution of the witness beam in the first (a) and last (b) interstage section.
Dashed vertical lines indicate the active plasma lens. 
Dashed orange lines indicate the dipoles composing the chicane.
}
\label{fig:3}
\end{figure}

 Considering the length of the dipoles as available drifts for the optics (i.e., the dipoles occupy the whole distance available between plasma stages and focusing elements), it is possible to capture and re-focus the beam using active plasma lenses (APL)~\cite{CHEN:1987,VAN_TILBORG:2015}.
For the parameters under consideration, the APL at the first interstage section ($E_W=27\,$GeV) should be 13 cm long, have a radius of 1 mm, and operate with a current pulse of amplitude I=3kA. [see Fig.~\ref{fig:3}(a)].
Along the linac, the length of the optical device increases by $\propto\sqrt{\gamma}$, for the same current pulse. 
Therefore, the length of the dipoles can be increased, and the magnetic field tapered (as $\propto 1/\gamma$), further decreasing the amount of emitted radiation.  
This is also useful to mitigate the degrading of the transverse beam quality due to the increase of energy spread in the dispersive sections~\cite{RAUBENHEIMER:1993,LINDSTROM:2025}.
Figure~\ref{fig:3}(b) shows the scheme of the last interstage section ($E_W=4.982\,$TeV).

\par
\begin{figure}[!h]
\centering
\includegraphics[width=\columnwidth]{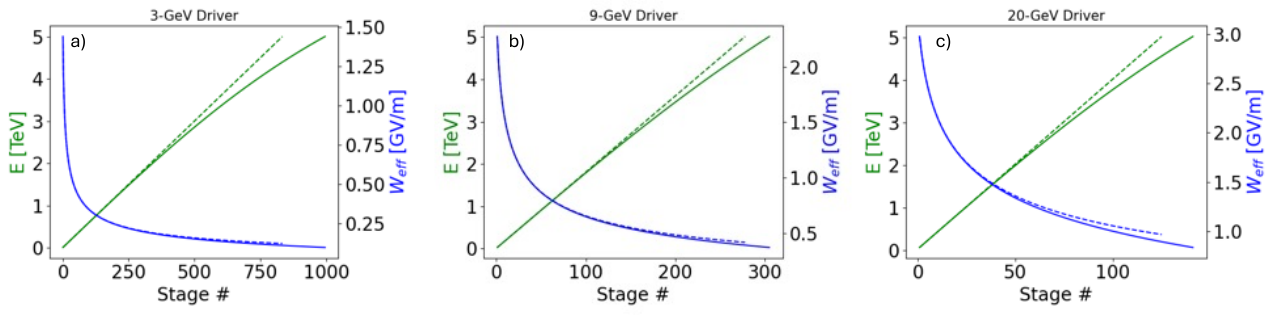} 
\caption{Energy of the witness bunch (green line, left-hand side vertical axis) and effective gradient (blue line, right-hand side axis vertical axis) for 3-GeV (a), 9-GeV (b), 20-GeV (c) driver bunches.}
\label{fig:4}
\end{figure}

 Figure~\ref{fig:4} shows the effective gradient (blue lines, left-hand side vertical axis) for different energies of the electron in the driver, with active plasma lenses and constant (continuous lines) or tapered (dashed lines) magnetic dipoles.
Calculations show that, for $E_D$ and initial $E_W\gtrsim 9\,$GeV, it is possible to achieve an effective gradient $>0.5\,$GV/m, and that the tapered configuration allows for a reduction of the number of stages needed to achieve the target energy.
For a simpler comparison, we have assumed the same $W_z^+$ for all cases, but we note that it could be increased with the energy of the driver bunches since they could be focused more tightly, increasing the charge density and the wakefield amplitude.

\par The effect of energy gain fluctuations (both from event to event and between successive stages) must be accounted for in the design of a PWFA linac. 
Such variations can originate from jitters in the plasma electron density, as well as of beam-to-beam alingment and synchronization. 
Proof-of-principle experiments have demonstrated that energy gain stability at the level of $\ll1\%$ is achievable~\cite{POMPILI:2021,LINDSTROM:2021a}.
Moreover, for the technical design of a real machine, it will be important to take into account the amount of power emitted as radiation, for safety and for machine protection. 

\section{Conclusions}

We discussed the effect of synchrotron radiation on the effective accelerating gradient of a staged, beam-driven, plasma wakefield accelerator. 
We showed that, to achieve the target energy of 5\,TeV, it is necessary to reduce the amplitude of the magnetic field in the dipoles separating in energy the spent driver from the witness bunch, so that less energy is lost due to the emission of synchrotron radiation. 
Additionally, effective accelerating gradients above 0.5\,GV/m can be achieved with the use of active plasma lenses as optical elements to transport and match the witness beam from one plasma stage to the next. 

\par We note that, while the use of higher-energy driver bunches could extend the length of each plasma stage and increase the net energy gain per stage (thereby reducing the total number of stages and potentially improving the effective gradient), these benefits must be weighed against the increased complexity and cost of a longer injector chain. 
Such requirements may offset the inherent advantages of plasma-based acceleration. 
Therefore, a careful optimization of parameters will be essential to determine the most effective configuration, taking into account the required energy reach and the overall machine footprint.

\bmhead{Acknowledgements}
We gratefully acknowledge useful discussions with the 10\,TeV Wakefield Collider Design Study Collaboration and with the PWFA Working Group. 
No new data were generated or analyzed in this study.

\bibliography{sn-bibliography}

@book{JACKSON:1962,
  title={Classical Electrodynamics},
  author={Jackson, John David},
  year={1962},
  publisher={John Wiley \& Sons}, 
address ={New York, USA}
}

@article{CHEN:1987,
	

	author = {Chen, P. },

	journal = {Particle Accelerators},
	number = {3-4},
	pages = {171-182},
	title = {A possible final focusing mechanism for linear colliders},
	url = {http://inis.iaea.org/search/search.aspx?orig_q=RN:18041971},
	volume = {20},
	year = {1987}
}

@ARTICLE{CHEN:1985,
  title = {Acceleration of Electrons by the Interaction of a Bunched Electron Beam with a Plasma},
  author = {Chen, Pisin and Dawson, J. M. and Huff, Robert W. and Katsouleas, T.},
  journal = {Physical Review Letters},
  volume = {54},
  issue = {7},
  pages = {693--696},
  numpages = {0},
  year = {1985},
  month = {Feb},
  publisher = {American Physical Society},
  doi = {10.1103/PhysRevLett.54.693},
  url = {https://link.aps.org/doi/10.1103/PhysRevLett.54.693}
}

@article{VERRA:2022b,
  title = {Controlled Growth of the Self-Modulation of a Relativistic Proton Bunch in Plasma},
  author = {{Verra, L. et al. (AWAKE Coll.)}},
  collaboration = {AWAKE Collaboration},
  journal = {Phys. Rev. Lett.},
  volume = {129},
  issue = {2},
  pages = {024802},
  numpages = {7},
  year = {2022},
  month = {Jul},
  publisher = {American Physical Society},
  doi = {10.1103/PhysRevLett.129.024802},
  url = {https://link.aps.org/doi/10.1103/PhysRevLett.129.024802}
}

@ARTICLE{TAJIMA:1979,
title = {Laser Electron Accelerator},
  author = {Tajima, T. and Dawson, J. M.},
  journal = {Physical Review Letters},
  volume = {43},
  issue = {4},
  pages = {267--270},
  numpages = {0},
  year = {1979},
  month = {Jul},
  publisher = {American Physical Society},
  doi = {10.1103/PhysRevLett.43.267},
  url = {https://link.aps.org/doi/10.1103/PhysRevLett.43.267}
}

@article{DAWSON:1959,
  title = {Nonlinear Electron Oscillations in a Cold Plasma},
  author = {Dawson, John M.},
  journal = {Phys. Rev.},
  volume = {113},
  issue = {2},
  pages = {383--387},
  numpages = {0},
  year = {1959},
  month = {Jan},
  publisher = {American Physical Society},
  doi = {10.1103/PhysRev.113.383},
  url = {https://link.aps.org/doi/10.1103/PhysRev.113.383}
}

@ARTICLE{BATSCH:2021,
	AUTHOR = "{{F. Batsch, P. Muggli et al. (AWAKE Coll.)}}",
	title = {Transition between Instability and Seeded Self-Modulation of a Relativistic Particle Bunch in Plasma},
  collaboration = {AWAKE Collaboration},
  journal = {Physical Review Letters},
  volume = {126},
  issue = {16},
  pages = {164802},
  numpages = {6},
  year = {2021},
  month = {Apr},
  publisher = {American Physical Society},
  doi = {10.1103/PhysRevLett.126.164802},
  url = {https://link.aps.org/doi/10.1103/PhysRevLett.126.164802}
}

@ARTICLE{AWAKE:2018,
	AUTHOR = "{{AWAKE Collaboration}}",
	TITLE = " {Acceleration of electrons in the plasma wakefield of a proton bunch}",
	JOURNAL = "Nature ",
	YEAR="2018",
    volume = {561},
    number = {7723},
    pages = {363-367},
    doi = {10.1038/s41586-018-0485-4},
    URL = {https://doi.org/10.1038/s41586-018-0485-4},
}

@ARTICLE{KUMAR:2010,
  title = {Self-Modulation Instability of a Long Proton Bunch in Plasmas},
  author = {Kumar, Naveen and Pukhov, Alexander and Lotov, Konstantin},
  journal = {Physical Review Letters},
  volume = {104},
  issue = {25},
  pages = {255003},
  numpages = {4},
  year = {2010},
  month = {Jun},
  publisher = {American Physical Society},
  doi = {10.1103/PhysRevLett.104.255003},
  url = {https://link.aps.org/doi/10.1103/PhysRevLett.104.255003}
}

@article{CHEN:1986,
  title = {Energy Transfer in the Plasma Wake-Field Accelerator},
  author = {Chen, Pisin and Su, J. J. and Dawson, J. M. and Bane, K. L. F. and Wilson, P. B.},
  journal = {Phys. Rev. Lett.},
  volume = {56},
  issue = {12},
  pages = {1252--1255},
  numpages = {0},
  year = {1986},
  month = {Mar},
  publisher = {American Physical Society},
  doi = {10.1103/PhysRevLett.56.1252},
  url = {https://link.aps.org/doi/10.1103/PhysRevLett.56.1252}
}

@article{POMPILI:2021,
	author = {Pompili, R. and Alesini, D. and Anania, M. P. and Behtouei, M. and Bellaveglia, M. and Biagioni, A. and Bisesto, F. G. and Cesarini, M. and Chiadroni, E. and Cianchi, A. and Costa, G. and Croia, M. and Del Dotto, A. and Di Giovenale, D. and Diomede, M. and Dipace, F. and Ferrario, M. and Giribono, A. and Lollo, V. and Magnisi, L. and Marongiu, M. and Mostacci, A. and Piersanti, L. and Di Pirro, G. and Romeo, S. and Rossi, A. R. and Scifo, J. and Shpakov, V. and Vaccarezza, C. and Villa, F. and Zigler, A.},
	date = {2021/04/01},
	doi = {10.1038/s41567-020-01116-9},
	id = {Pompili2021},
	isbn = {1745-2481},
	journal = {Nature Physics},
	number = {4},
	pages = {499--503},
	title = {Energy spread minimization in a beam-driven plasma wakefield accelerator},
	url = {https://doi.org/10.1038/s41567-020-01116-9},
	volume = {17},
	year = {2021}}

@article{LINDSTROM:2021a,
  title = {Energy-Spread Preservation and High Efficiency in a Plasma-Wakefield Accelerator},
  author = {Lindstr\o{}m, C. A. and Garland, J. M. and Schr\"oder, S. and Boulton, L. and Boyle, G. and Chappell, J. and D'Arcy, R. and Gonzalez, P. and Knetsch, A. and Libov, V. and Loisch, G. and Martinez de la Ossa, A. and Niknejadi, P. and P\~oder, K. and Schaper, L. and Schmidt, B. and Sheeran, B. and Wesch, S. and Wood, J. and Osterhoff, J.},
  journal = {Phys. Rev. Lett.},
  volume = {126},
  issue = {1},
  pages = {014801},
  numpages = {6},
  year = {2021},
  month = {Jan},
  publisher = {American Physical Society},
  doi = {10.1103/PhysRevLett.126.014801},
  url = {https://link.aps.org/doi/10.1103/PhysRevLett.126.014801}
}

@article{LITOS:2014,
	author = {{Litos, M. et al.}},
	date = {2014/11/01},
	doi = {10.1038/nature13882},
	id = {Litos2014},
	isbn = {1476-4687},
	journal = {Nature},
	number = {7525},
	pages = {92--95},
	title = {High-efficiency acceleration of an electron beam in a plasma wakefield accelerator},
	url = {https://doi.org/10.1038/nature13882},
	volume = {515},
	year = {2014}}

@misc{GESSNER:2025,
      title={Design Initiative for a 10 TeV pCM Wakefield Collider}, 
      author={{Gessner, S. et al.}},
      year={2025},
      eprint={2503.20214},
      archivePrefix={arXiv},
      primaryClass={physics.acc-ph},
      url={https://arxiv.org/abs/2503.20214}, 
}

@article{CALDWELL:2009,
	author = {Caldwell, Allen and Lotov, Konstantin and Pukhov, Alexander and Simon, Frank},
	date = {2009/05/01},
	doi = {10.1038/nphys1248},
	id = {Caldwell2009},
	isbn = {1745-2481},
	journal = {Nature Physics},
	number = {5},
	pages = {363--367},
	title = {Proton-driven plasma-wakefield acceleration},
	url = {https://doi.org/10.1038/nphys1248},
	volume = {5},
	year = {2009}}

@misc{CALDWELL:2025,
      title={Proton-Driven Plasma Wakefield Acceleration for Future HEP Colliders}, 
      author={{Caldwell, A. et al.}},
      year={2025},
      eprint={2503.21669},
      archivePrefix={arXiv},
      primaryClass={physics.acc-ph},
      url={https://arxiv.org/abs/2503.21669}, 
}

@article{LINDSTROM:2021b,
  title = {Staging of plasma-wakefield accelerators},
  author = {Lindstr\o{}m, Carl A.},
  journal = {Phys. Rev. Accel. Beams},
  volume = {24},
  issue = {1},
  pages = {014801},
  numpages = {20},
  year = {2021},
  month = {Jan},
  publisher = {American Physical Society},
  doi = {10.1103/PhysRevAccelBeams.24.014801},
  url = {https://link.aps.org/doi/10.1103/PhysRevAccelBeams.24.014801}
}

@article{BARKLOW:2023,
doi = {10.1088/1748-0221/18/09/P09022},
url = {https://doi.org/10.1088/1748-0221/18/09/P09022},
year = {2023},
month = {sep},
publisher = {IOP Publishing},
volume = {18},
number = {09},
pages = {P09022},
author = {{Barklow, T. et al.}},
title = {Beam delivery and beamstrahlung considerations for ultra-high energy linear colliders},
journal = {Journal of Instrumentation}
}

@article{BARKLOW:2023b,
doi = {10.1088/1748-0221/18/07/P07028},
url = {https://doi.org/10.1088/1748-0221/18/07/P07028},
year = {2023},
month = {jul},
publisher = {IOP Publishing},
volume = {18},
number = {07},
pages = {P07028},
author = {{Barklow, T. et al.}},
title = {XCC: an X-ray FEL-based γγ Compton collider Higgs factory},
journal = {Journal of Instrumentation}
}

@misc{LINDSTROM:2021c,
      title={Self-correcting longitudinal phase space in a multistage plasma accelerator}, 
      author={Carl A. Lindstrøm},
      year={2021},
      eprint={2104.14460},
      archivePrefix={arXiv},
      primaryClass={physics.acc-ph},
      url={https://arxiv.org/abs/2104.14460}, 
}

@article{FOSTER:2025,
title = {Proceedings of the Erice workshop: A new baseline for the hybrid, asymmetric, linear Higgs factory HALHF},
journal = {Physics Open},
volume = {23},
pages = {100261},
year = {2025},
issn = {2666-0326},
doi = {https://doi.org/10.1016/j.physo.2025.100261},
url = {https://www.sciencedirect.com/science/article/pii/S2666032625000110},
author = {{Foster, B. et al.}}
}

@article{VERRA:2020,
doi = {10.1088/1742-6596/1596/1/012007},
url = {https://dx.doi.org/10.1088/1742-6596/1596/1/012007},
year = {2020},
month = {jul},
publisher = {IOP Publishing},
volume = {1596},
number = {1},
pages = {012007},
author = {L. Verra and E. Gschwendtner and P. Muggli},
title = {{Study of external electron beam injection into proton driven plasma wakefields for AWAKE Run 2}},
journal = {Journal of Physics: Conference Series}
}

@article{MUGGLI:2004,
  title = {Meter-Scale Plasma-Wakefield Accelerator Driven by a Matched Electron Beam},
  author = {{Muggli, P. et al.}},
  journal = {Physical Review Letters},
  volume = {93},
  issue = {1},
  pages = {014802},
  numpages = {4},
  year = {2004},
  month = {Jun},
  publisher = {American Physical Society},
  doi = {10.1103/PhysRevLett.93.014802},
  url = {https://link.aps.org/doi/10.1103/PhysRevLett.93.014802}
}

@misc{Barnes:2010,
      author        = "Barnes, M.J. and Borburgh, J. and Goddard, B. and
                       Hourican, M.",
      title         = "{Injection and extraction magnets: septa}",
      archivePrefix = "arXiv",
      eprint        = "1103.1062",
      year          = "2010",
      url           = "https://cds.cern.ch/record/1334285",
      note          = "Comments: 18 pages, presented at the CERN Accelerator
                       School CAS 2009: Specialised Course on Magnets, Bruges,
                       16-25 June 2009",
      doi           = "10.5170/CERN-2010-004.167",
}

@article{VAN_TILBORG:2015,
  title = {Active Plasma Lensing for Relativistic Laser-Plasma-Accelerated Electron Beams},
  author = {{van Tilborg, J. et al.}},
  journal = {Phys. Rev. Lett.},
  volume = {115},
  issue = {18},
  pages = {184802},
  numpages = {5},
  year = {2015},
  month = {Oct},
  publisher = {American Physical Society},
  doi = {10.1103/PhysRevLett.115.184802},
  url = {https://link.aps.org/doi/10.1103/PhysRevLett.115.184802}
}

@inproceedings{RAUBENHEIMER:1993,
  author       = {{Raubenheimer, T., Emma, P., Kheifets, S.}},
  title        = {{Chicane and Wiggler Based Bunch Compressors for Future Linear Colliders}},
  booktitle    = {Proc. PAC'93},
  year         = {1993},
}

@misc{LINDSTROM:2025,
      title={Achromatic optics using nonlinear plasma lenses for beam-quality preservation between plasma-accelerator stages}, 
      author={C. A. Lindstrøm and E. Adli and J. B. B. Chen and P. Drobniak and A. Huebl and D. Kalvik and C. E. Mitchell and F. Peña and K. N. Sjobak},
      year={2026},
      eprint={2604.17605},
      archivePrefix={arXiv},
      primaryClass={physics.acc-ph},
      url={https://arxiv.org/abs/2604.17605}, 
}

@misc{LCVISION:2025,
      title={A Linear Collider Vision for the Future of Particle Physics}, 
      author={{H. Abramowicz et al.}},
      year={2025},
      eprint={2503.19983},
      archivePrefix={arXiv},
      primaryClass={hep-ex},
      url={https://arxiv.org/abs/2503.19983}, 
}

@inproceedings{DIMITRI:2016,
  author       = {{S. Di Mitri}},
  title        = {{Bunch-length Compressors}},
  booktitle    = {Proceedings of the CAS–CERN Accelerator School: Free Electron Lasers and Energy Recovery Linacs, Hamburg, Germany, 31 May–10 June 2016, CERN Yellow Reports},
  year         = {2016},
}

@article{ZHANG:2025,
	author = {Zhang, Chaojie and Storey, Douglas and Knetsch, Alexander and O'Shea, Brendan D. and Ariniello, Robert and Cao, Gevy J. and Corde, S{\'e}bastien and Dalichaouch, Thamine N. and Emma, Claudio and Finnerud, Ole G. and Gessner, Spencer and Hansel, Claire and Hansen, Elias and Lee, Valentina and Lindstr{\o}m, Carl A. and Litos, Michael and Majernik, Nathan and Marsh, Kenneth A. and Mori, Warren B. and Rajkovic, Ivan and Hogan, Mark J. and Joshi, Chan},
	date = {2025/11/28},
	doi = {10.1038/s41467-025-65742-8},
	id = {Zhang2025},
	isbn = {2041-1723},
	journal = {Nature Communications},
	number = {1},
	pages = {10719},
	title = {Plasma-wakefield accelerator simultaneously boosts electron beam energy and brightness},
	url = {https://doi.org/10.1038/s41467-025-65742-8},
	volume = {16},
	year = {2025}}

\end{document}